\documentclass[preprint,showpacs,preprintnumbers,superscriptaddress,amsmath,amssymb]{revtex4}
 \usepackage{cases}
 \usepackage{CJK}
 \usepackage{txfonts}
 \usepackage{mathrsfs}
 \usepackage{amssymb}
 \usepackage[dvips]{graphicx}% Include figure files
 \usepackage{dcolumn}% Align table columns on decimal point
 \usepackage{bm}% bold math
 \usepackage{subfigure}
 \usepackage{floatflt}
 \usepackage{float}
 \usepackage{rotating}
 \usepackage{multirow}
 \usepackage[bookmarksnumbered,bookmarksopen,colorlinks,citecolor=blue,linkcolor=blue]{hyperref} %
 \usepackage{epsfig,graphics,psfrag,amsmath}
 \usepackage{sidecap}
\def\gxnu{College of Physics and Technology, Guangxi Normal University, Guilin 541004, China}
\def\gxnp{Guangxi Key Laboratory of Nuclear Physics and Technology, Guangxi Normal University, Guilin 541004, China}

\def\F17e{$^{17}$F$^*$}

\begin{document}
%\begin{CJK*}{GBK}{song}

 %\preprint{00-000}

 \title{Shielding effects in fusion reactions of proton-halo nucleus}

 \author{Xue-ying He}
 \affiliation{\gxnu}\affiliation{\gxnp}
 \author{Qin Dong}
 \affiliation{\gxnu}\affiliation{\gxnp}

 \author{Li Ou}\email{liou@gxnu.edu.cn}
 \affiliation{\gxnu}\affiliation{\gxnp}

 \date{\today}% It is always \today, today,

\begin{abstract}
  To explain the experimental facts that the fusion cross sections of proton-halo nucleus on heavy target nucleus is not enhanced as expected, the shielding supposition has been proposed. Namely, the proton-halo nucleus is polarized with the valence proton being shielded by the core. In this paper, within the frame of the Improved Quantum Molecular Dynamics model, the fusion reactions by $^{17}$F on $^{208}$Pb around Coulomb barrier have been simulated. The existence of shielding effect is verified by microscopic dynamics analysis and its influence on the effective interaction potential is also investigated.
\end{abstract}


 \maketitle

\section{Introduction}

 The heavy ion fusion reaction has attracted much attention in recent years \cite{hofmann00,oganessian10,adamian00,lu12}, because it is not only the one of important ways to synthesize new super-heavy-element (SHE) but also an important method to study the nuclear structure and nuclear reaction mechanism. With the rare radioactive beams provided by many laboratories around the world, one can further study properties of nuclei located nearly the drip lines and study the nuclear reactions by using projectiles with exotic properties drastically different from those along $\beta$-stable line. Because of new phenomena in weakly bounded nuclei \cite{liang00,liu04,takigawa93,hussein93,dasso96,hagino00,vinodkuma13,dasso92}, such as increased probabilities for neutron transfer and fusion \cite{signorini97}, increased breakup probability, the one of hot topics in SHE synthesis is the fusion with weakly bounded nucleus, especially the halo nucleus.

 In particular, halo nucleus, in which the valence neutron(s) or proton(s) are very weakly bound, are expected to give rise to new dynamics phenomena in nuclear reactions. In a semiclassical picture, the nucleus with low binding energy usually has larger radius. So the probabilities for specific reaction channels such as neutron transfer and fusion will increase \cite{signorini97}. But the experiments of the proton drip line nucleus $^{17}$F on $^{208}$Pb at energies around the Coulomb barrier do not observe the enhancement of the fusion cross section due to breakup or to a large interaction radius\cite{rehm98}. With experiment on reaction of $^{17}$F+$^{208}$Pb at energies around Coulomb barrier, Liang proved that the influence of breakup of $^{17}$F on fusion cross sections is quite weak \cite{liang00}. One of suppositions to explain the experimental data is so called the ``shielding effect'': During the reaction, $^{17}$F is excited into the first excited state to be a proton-halo nucleus (hereinafter denoted as \F17e), and is polarized by the valence proton being repelled from the reaction partner and being shielded by the $^{16}$O core. With such a shielding effect, the effective interaction radius is smaller than the one with $^{17}$F in ground state, and the fusion rate will not rise as expected.

 In this work, with the improved quantum molecular dynamics (ImQMD) model, taking $^{17}$F+ $^{208}$Pb as example, we try to verify existence of the shielding effect of valence proton in the fusion reactions and investigate the influence of shielding effect on the dynamic effective interaction potential. The paper is organized as follows.
 In Sec. 2, we briefly introduce the model we adopted.
 In Sec. 3, we present the improvement on the ImQMD model to describe the halo nucleus and the shielding effect on fusion reactions.  Finally a brief summary is given in Sec. 4.

\section{Model}

 The quantum molecular dynamics (QMD) model \cite{aichelin88,aichelin91} built by  Aichelin and Peilert et. al., which can intuitively present the microscopic dynamics process of nuclear reaction, is a powerful tool to study the heavy ion collisions (HICs) at intermediate energies. Several versions based on QMD model, such as ImQMD, IQMD, IQMD-BNU, IQMD-IMP, IQMD-SINAP, JAM, JQMD, TuQMD, UrQMD, et. al., have been developed and successfully applied in various research fields.
 The detailed discussion of the QMD models can be found in Refs \cite{xu19,zhang18,xu16}. The ImQMD model \cite{wang02,wang04,zhang05,ou08,ou15,tian08,zhao13} is used in this work, which is briefly introduced hereby.

 In the ImQMD model, the nucleon is represented by a Gaussian wave-packet, which reads
 \begin{eqnarray}\label{phi}
   \phi_{i}(\bm{r})=\frac{1}{(2\pi\sigma_{r}^{2})^{3/4}}
   \exp\left[-\frac{(\bm{r}-\bm{r}_{i})^{2}}{4\sigma_{r}^{2}}
   +\frac{i}{\hbar}\bm{r}\cdot\bm{p}_{i}\right],
 \end{eqnarray}
 where $\bm{r}_{i}$, $\bm{p}_{i}$ is the center of wave-packet for the $i$th nucleon in coordinate and momentum space, respectively. And $\sigma_{r}$ is the width of wave packet in coordinate space. With system size effect considered, $\sigma_r$ can be expressed as
 \begin{eqnarray}\label{sigma}
 \sigma_{r}=\sigma_{0}A^{1/3}+\sigma_{1},
 \end{eqnarray}
 where $A$ is nucleon number of nucleus, $\sigma_{0}$ and $\sigma_{1}$ are parameters obtained by fitting the properties of many $\beta$-stable nuclei. By making the Wigner transform on the nucleon wave function and summing all nucleons up, one can get the phase space distribution function for nuclear system, which reads
   \begin{eqnarray}\label{frp}
 f(\bm{r},\bm{p})=\sum _{i}^{A} f_{i}(\bm{r},\bm{p})=\sum_{i=1}^{A}\frac{1}{(\pi\hbar)^3}
 \exp\left[-\frac{(\bm{r}-\bm{r}_i)^2}{2\sigma_{r}^{2}}\right]
 \exp\left[-\frac{(\bm{p}-\bm{p}_i)^2}{2\sigma_{p}^{2}}\right],
 \end{eqnarray}
 where $\sigma_{p}$ is the width of wave packet in momentum space, which satisfies the minimum uncertainty relation $\sigma_r \cdot \sigma_p = \textstyle{\hbar \over 2}$. The center of wave-packet for each nucleon evolves following Hamilton canonical equation, \begin{eqnarray}\label{HCE}
 \dot{\bm{r}}_{i}=\frac{\partial H}{\partial\bm{p}_{i}},~~~~~~
 \dot{\bm{p}}_{i}=-\frac{\partial H}{\partial\bm{r}_{i}}.
 \end{eqnarray}
 The Hamiltonian, including kinetic energy, Coulomb energy and nuclear local potential energy, reads
 \begin{eqnarray}\label{H}
 H=T+U_{\rm{Coul}}+U_{\rm{loc}}.
 \end{eqnarray}
 Each term of Hamiltonian is written as
 \begin{eqnarray}\label{T}
 T=\sum_{i}^{A}\frac{p_i^2}{2m},
 \end{eqnarray}
 \begin{eqnarray}\label{Ucoul}
 U_{\rm Coul}=\frac{1}{2}\int\rho_p({\bm r})\frac{e^2}{|{{\bm r}-{{\bm r}'}}|}\rho_p({\bm r}')d{\bm r}d{{\bm r}'},
 \end{eqnarray}
 \begin{eqnarray}\label{Uloc}
 U_{\rm loc}=\int V_{\rm loc}d{\bm r},
 \end{eqnarray}
 where $V_{\rm{loc}}$ is the Skyrme potential energy density functional without the spin-orbit term considered, which can be expressed as
 \begin{eqnarray}\label{Vloc}
 V_{\rm{loc}}=\frac{\alpha}{2}\frac{\rho^{2}}{\rho_{0}}+
 \frac{\beta}{\gamma+1}\frac{\rho^{\gamma+1}}{\rho_{0}^{\gamma}}+
 \frac{g_{\rm sur}}{2}(\bigtriangledown\rho)^{2}
 +g_{\tau}\frac{\rho^{\eta+1}}{\rho_{0}^{\eta}}+
 \frac{C_{\rm s}}{2\rho_{0}}[\rho^{2}-\kappa_{\rm s}(\bigtriangledown\rho)^{2}]\delta^{2}.
 \end{eqnarray}
 Here $\rho$ and $\rho_0$ are the nucleon density and the saturation density, respectively. And $\delta = (\rho_n -\rho_p)/\rho$ is the isospin asymmetry degree. The parameter set used in Eq. (\ref{Vloc}), named IQ3a which has been successfully applied to study the fusion reactions \cite{wang14,wang15}, is listed in Table \ref{IQ3a}.
  \begin{table*}[htbp]
 \caption{\label{IQ3a}The model parameter set IQ3a.}
 \begin{center}
 \begin{tabular}{ccccccccccc}
 \hline\hline
 $\alpha $ & $\beta $ & $\gamma $ &$%
 g_{\rm{sur}}$ & $ g_{\tau }$ & $\eta $ & $C_{\rm{s}}$ & $\kappa _{\rm{s}}$ &
 $\rho_{0}$ & ~~$\sigma_0$~~ & ~~$\sigma_1$~~ \\
 (MeV) & (MeV) & & (MeV~fm$^{2}$) & (MeV) & & (MeV) & (fm$^{2}$) &
 (fm$^{-3}$) & (fm) & (fm) \\ \hline
 $-207$ & 138 & 7/6 & 16.5 & 14.0 & 5/3 & 34.0 & 0.4 & 0.165 & 0.02 & 0.94\\
 \hline\hline
 \end{tabular}
 \end{center}
 \end{table*}

 \section{Results and discussion}

 \subsection{Preparation of halo nucleus}

 The first step to simulate the reaction by ImQMD model is initialization of projectile and target nuclei. As is well known, the good initial nucleus is very important for QMD calculations, especially for the reactions at low energies. The typical method to the prepare initial nuclei in QMD is as the following: First, the position of each nucleon is determined by Monte carlo sampling according to the size of nucleus. Second, based on the density distribution
 \begin{eqnarray}\label{rhor}
 \rho(\bm{r})=\int{f(\bm{r},\bm{p})}d{\bm p}=\frac{1}{(2\pi\sigma_{r}^{2})^{3/2}}
 \sum_{i=1}^{A}\exp\left[-\frac{(\bm{r}-\bm{r}_{i})^{2}}{2\sigma_{r}^{2}}\right],
 \end{eqnarray}
 the local Fermi momentum is obtained by the local-density approximation $P_F=(\textstyle{3\pi^2 \over 2}\rho)^{1/3}-\Delta p_F$, here $\Delta P_F$ is the contribution from the width of wave packet and determined by binding energy. Then the initial momentum of each nucleon is sampled randomly in range of 0-$P_F$. To check the stability of the prepared nucleus, the prepared nuclear system solely evolves for a period of time short or long depending on the type of reaction. Only those prepared nuclei with good properties, such as the root-mean-square (rms) radii, the binding energies, and their time evolutions are good enough, and there is no spurious particle emission for a long enough time, are selected as ``good initial nuclei''.

 In order to describe the halo-nucleus, which is a weakly bound nucleus, by such a semi-classical model, the special treatment on halo-nucleus in ImQMD model must be adopted. The halo-nucleus is usually considered as a system with a loose valence nucleon moving around a tight core. Taking a proton-halo nucleus \F17e~for example, the rms radius of \F17e~is $2.71\pm0.18$ fm, and rms radius of valence proton is 5.33 fm \cite{morlock97,zhangh02,tilley93}. So \F17e~is considered as a system with $^{16}$O core and a valence proton. To save the CPU time, in ImQMD calculation, the initial distance between the projectile and target is not infinity as the real situation but a finite value of 40 fm.  There is a very large probability for $^{17}$F to be excited by the heavy target nucleus in a so short distance. So we sample \F17e~nucleus for initial state for ImQMD calculations. In ImQMD model, the $^{16}$O core is prepared by the conventional method. According to the experimental data, the rms radius of neutron of \F17e~in the first excited state is 2.478 fm, it means that $^{16}$O core of \F17e~is smaller than ordinary $^{16}$O nucleus with rms radius of matter of 2.710 fm. So the radius and the wave-packet width calculated based on ordinary $^{16}$O are both multiplied by 0.9 for $^{16}$O core sampling. In order to set valence proton with reasonable properties, two factors should be considered. The one is the position of valence proton which should be away from core to form the halo. The other is the wave-packet width of valence proton, which reflects the range of nuclear force, has obvious effect on nucleus surface diffusion \cite{maruyama96}. It is obvious that the wave-packet width expression (\ref{sigma}) for ordinary nucleon is not suitable for the loosely bound valence proton. The extreme surface-diffusion-structure of \F17e, i.e. halo, require the valence proton to have a relatively larger width of wave-packet. By optimally fitting the experimental data of \F17e, in the calculations, the initial distance between the valence proton and the center of mass of core is set to $R=R_{\rm p}+1.6$ fm, where $R_{\rm p}$ is the radius of proton distribution of $^{16}$O core \cite{pomorska94,myers95}, the wave-packet width and the initial momentum is set to $\sqrt{2}$ fm and 25 MeV/c, respectively, for the valence proton.

 The ImQMD model is a semiclassical model with Skyrme $\delta$-type interaction adopted. It is quite difficult for a weakly bound nucleus, such as halo nucleus, to keep stable for a long time in these kinds of model. The valence proton will finally fall into the core due to the strong attractiveness. To maintain the stable halo, a so called ``sphere-shell constrain'' is introduced into the model. When the valence proton is too close to the core and trends to move further toward the core, or too far from the core and trends to move backward the core, its momentum directions will be changed randomly by elastic scattering with nucleon nearby.

\begin{figure}[h]
 \centering
   \includegraphics[width=0.8\textwidth]{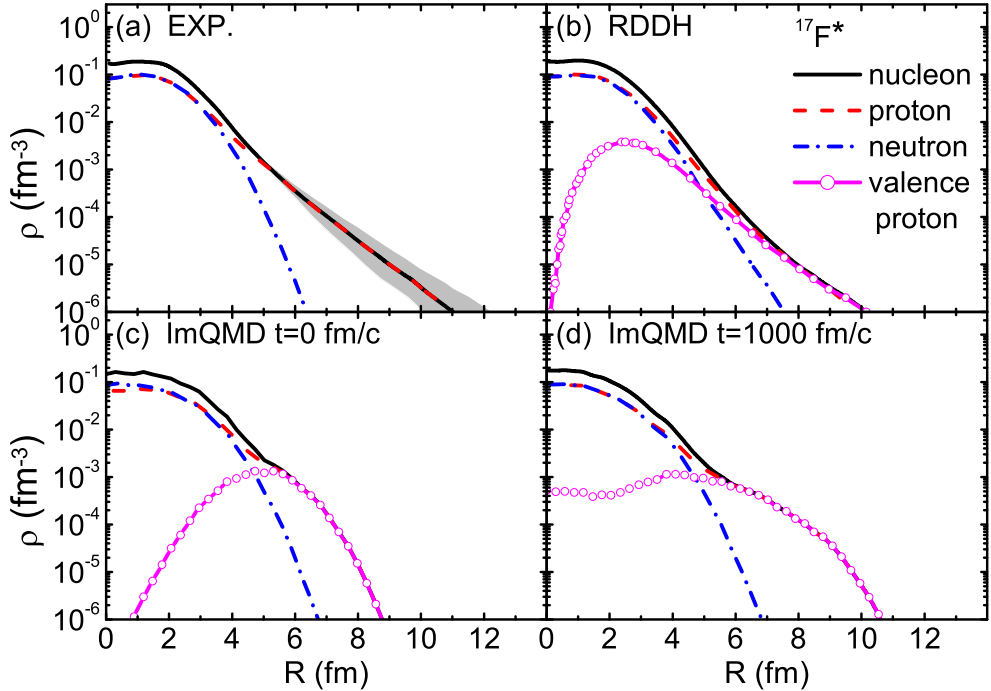}
   \caption{\label{fig1}(color online) Density distributions of nucleon, proton, neutron and the valence proton in \F17e~from experiment (a), relativistic density-dependent Hartree theory (b) and ImQMD calculations (c) (d).}
   \end{figure}

 By the modification on the preparation of initial nucleus and the introduction of sphere-shell constrain, we can get \F17e~with good properties and stable enough. In figure \ref{fig1}, the density distributions of nucleon, proton, neutron and the valence proton in \F17e~at 0 and 1000 fm/c calculated by ImQMD model are presented. The data taking from experiment \cite{zhangh02} and relativistic density-dependent Hartree theory (RDDH) \cite{zhangh02,ren98} are also shown for comparison. One can see that, although the halo of initial \F17e~is a little small, the distributions given by ImQMD model are reasonable. And by the nucleons self-adjustment with time evolution, ImQMD model gives a quite good distribution close to the results from experiments and RDDH theory.
   
\begin{figure}[h]
 \centering
   \includegraphics[width=0.8\textwidth]{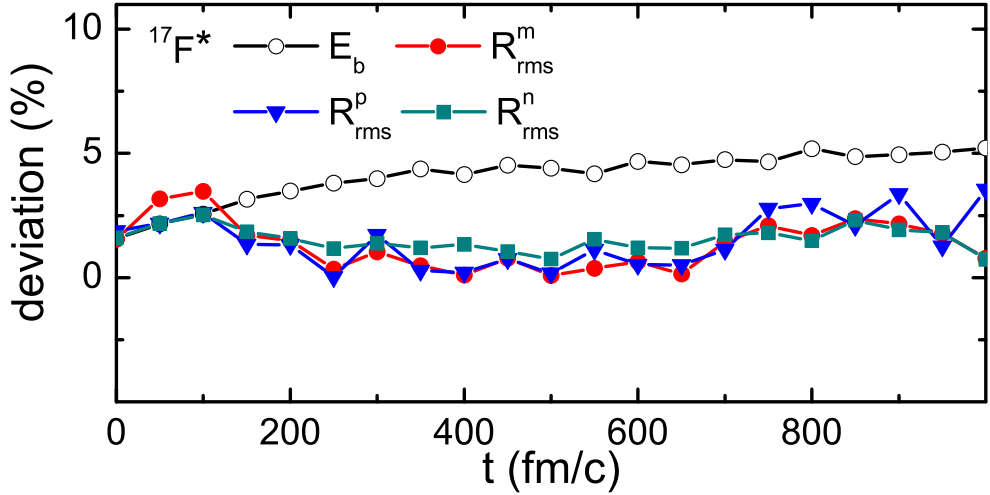}
   \caption{\label{fig2}(color online) Time evolutions of deviation of the binding energies and the root-mean-square radii of nucleon, proton and neutron for \F17e.}
   \end{figure}

    In figure \ref{fig2} we show the time evolutions of deviation of the binding energies and the root-mean-square radii of nucleon, proton, and neutron for \F17e. One can see that the binding energies and each root-mean-square radii remain stable with small fluctuation for a long enough time.
   
 \subsection{Proton-halo nucleus fusion reaction}
\begin{figure}[h]
 \centering
   \includegraphics[width=0.8\textwidth]{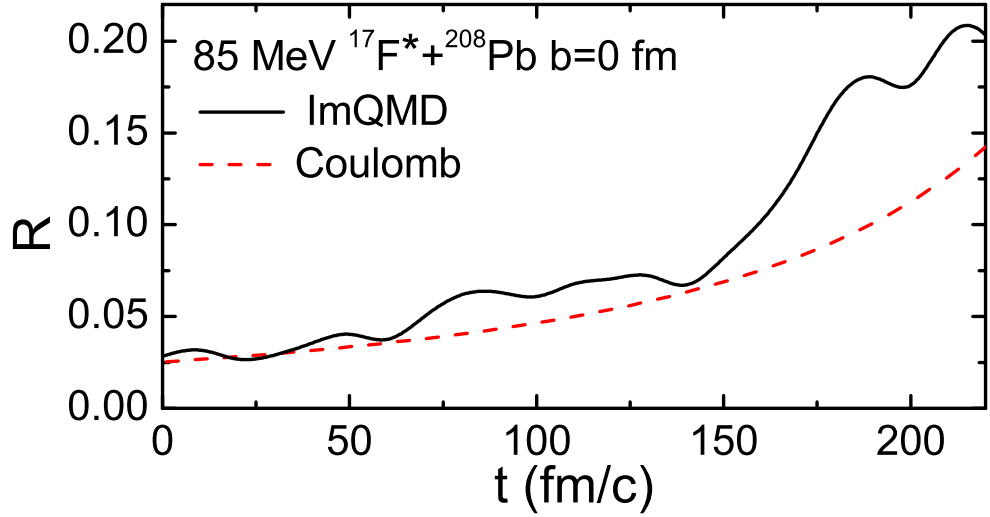}
   \caption{\label{fig3}(color online) Time evolution of ratio between force on valence proton from $^{208}$Pb and core of \F17e.}
   \end{figure}

 To obtain stable halo structure, ``sphere-shell constrain'', which is not a self-consistent treatment, is introduced into the model. The halo structure is easily destroyed when halo nucleus enter the field of target nucleus. So ``sphere-shell constrain'' should be switched off at a proper moment during the reaction simulation, then the valence proton can move self-consistently in the system field. To select a proper switch-time, the time evolution of ratio between the force on valence proton from $^{208}$Pb and core $^{16}$O is investigated, which is shown in figure \ref{fig3} with the solid curve. For a comparison, we also calculated the ratio between the Coulomb force on valence proton from $^{208}$Pb and core $^{16}$O, denoted by dashed curve, assuming that the projectile and target are two charged points located at their individual centers of mass and move with their initial velocities only. One can see that, at the early stage, the effect from target is quite weak and the Coulomb force dominates the force from target. The influence of target becomes strong rapidly from 150 fm/c, and the nuclear force influence becomes obvious. So the ``sphere-shell constrain'' is switched off at 150 fm/c in the following calculations.

\begin{figure}[h]
 \centering
   \includegraphics[width=0.8\textwidth]{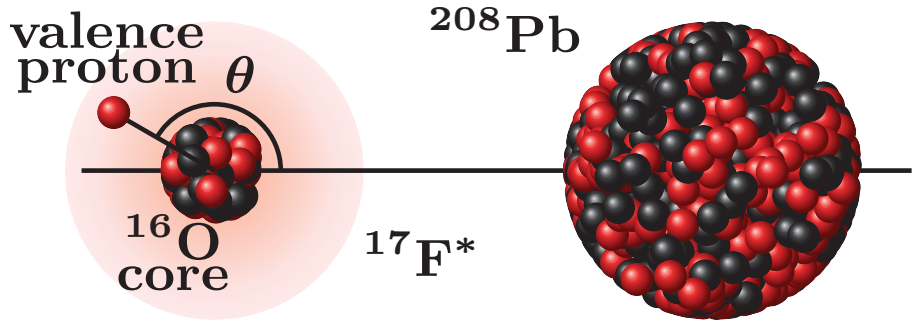}
   \caption{\label{fig4}(color online) Sketch for valence proton being shielded by core in reactions.}
   \end{figure}

\begin{figure}[h]
 \centering
   \includegraphics[width=0.8\textwidth]{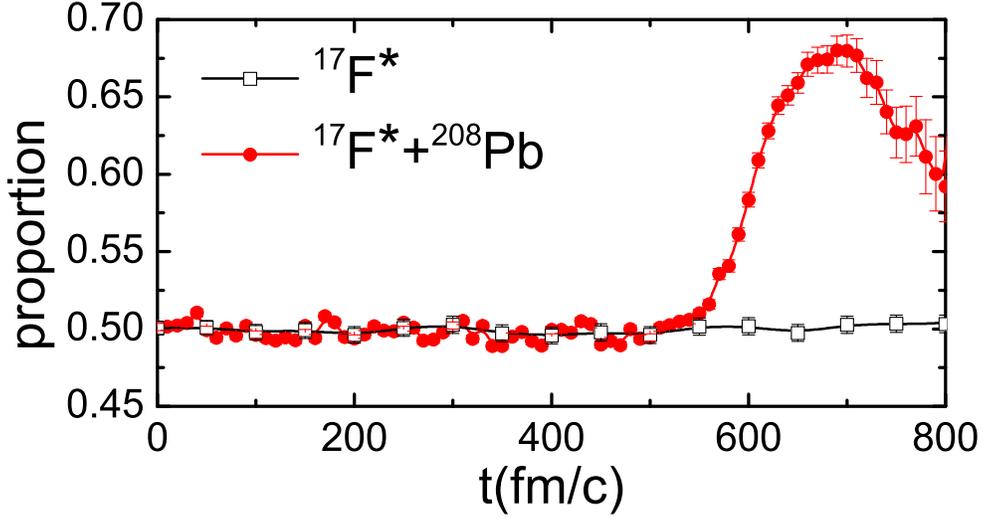}
   \caption{\label{fig5}(color online) Time evolution of proportion of events in which valence proton being shielded by core.}
   \end{figure}

    Firstly, we checked the shielding of valence proton by $^{16}$O core in reactions of \F17e+$^{208}$Pb at the energy $E_{\rm c.m.}=$ 85 MeV and $b=0$ fm. As illustrated in figure \ref{fig4}, the angle $\theta$ between the position vector of valence proton and the one of target's mass center from mass center of $^{16}$O can be a criterion for shielding. If $\cos \theta <0$, valence proton is considered as being shielded by $^{16}$O core.
   The proportion of the valence proton being shielded is defined as the ratio of $N_{\rm{s}}/N_{\rm{t}}$, which is shown in figure \ref{fig5}. Where $N_{\rm{s}}$ and $N_{\rm{t}}$ are the numbers of events in which valence proton being shielded during the reactions and the number of total events, respectively. The case for individual \F17e~is also shown for reference. One can see that, for individual \F17e, the valence proton distributes uniformly in the halo. In the \F17e+$^{208}$Pb reactions, at early stage in which projectile far away from target, there is not obvious shielding effect. With projectile being closer and closer to target, the shielding proportion arises. There are more and more events with valence proton being shielded by core from 500 fm/c. And after 700 fm/c when the $^{16}$O is absorbed by target to form the compound nucleus or elastically scattered off, the shielding effect is disturbed.
   
   By creating a certain amount of events at each incident energy $E_{\rm c.m.}$ and each impact parameter $b$ and counting the number of fusion events, the fusion probability $g_{\rm fus}$($E_{\rm c.m.}, b$) for a certain fusion reaction can be obtained to calculate the corresponding fusion cross seciton with
   \begin{eqnarray}\label{sigmaf}
    \sigma_{\rm fus}(E_{\rm c.m.}) =2\pi \int g_{\rm fus}(E_{\rm c.m.}, b) b db
     \simeq  2\pi \sum g_{\rm fus}(E_{\rm c.m.}, b_i) b_i \Delta b
  \end{eqnarray}
    where $\Delta b$ = 1 fm. The event is counted as a fusion (capture) event if the center-to-center distance between the two nuclei is smaller than the nuclear radius of the compound nuclei. Here the quasi-fission probability is neglected for the considered reaction systems \cite{wang14}.

\begin{figure}[h]
 \centering
   \includegraphics[width=0.8\textwidth]{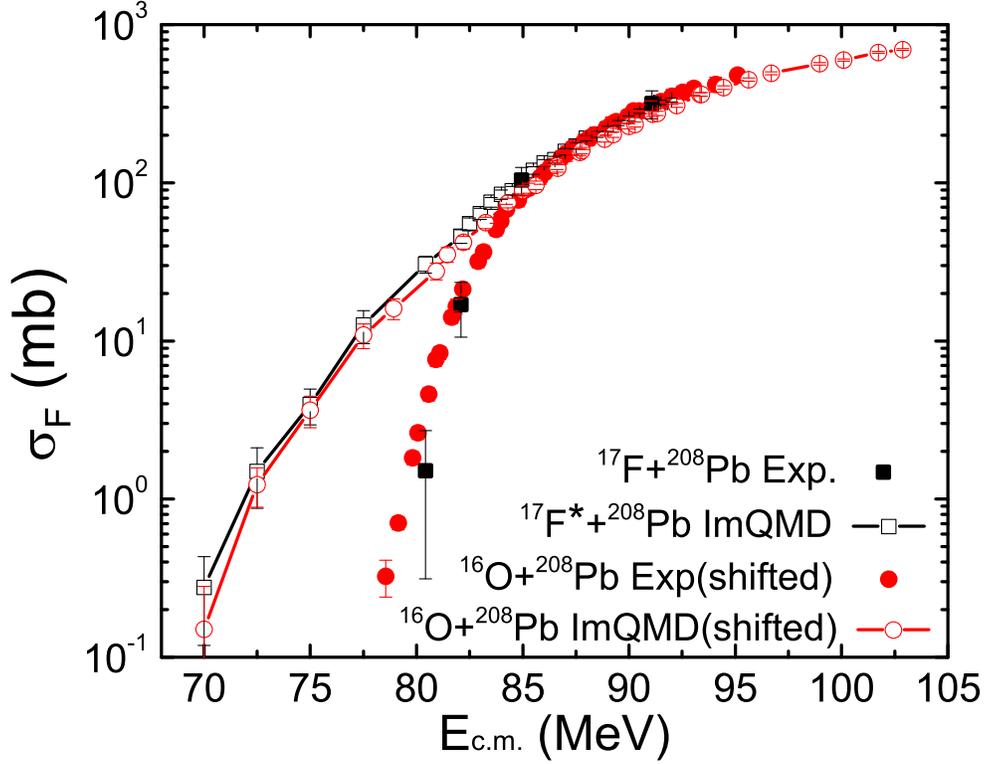}
   \caption{\label{fig6}(color online) Fusion excitation functions of \F17e+$^{208}$Pb and $^{16}$O+$^{208}$Pb calculated by ImQMD model compared to the experimental data. The results of $^{16}$O+$^{208}$Pb are shifted in energy by the factor 9/8.}
   \end{figure}

   In figure \ref{fig6}, we show the fusion excitation functions of \F17e+$^{208}$Pb and $^{16}$O+$^{208}$Pb calculated by ImQMD model. The experimental data are presented for comparison. The results of $^{16}$O+$^{208}$Pb are shifted in energy by factor 9/8 (the ratio of the charges of F and O). The fusion cross sections around and above Coulomb barrier are reproduced quite well; While for the case below Coulomb barrier, the fusion cross sections of \F17e+$^{208}$Pb is tiny higher than the ones of $^{16}$O+$^{208}$Pb, and they both overestimate the experimental data, which is probably attributable to the fact that the many quantum effects such as the shell effect and tunnelling effect can not be realized in the present semi-classical ImQMD model. Although the present version of the ImQMD model cannot describe the fusion of reactions at very low energies quite well, it is still reasonable to investigate the shielding effects by comparing the results of fusion cross sections between the \F17e+$^{208}$Pb and $^{16}$O+$^{208}$Pb. One can see that the fusion cross sections of \F17e+$^{208}$Pb and $^{16}$O+$^{208}$Pb are close to each other, with no enhancement observed whether for above or below Coulomb barrier.
   
\begin{figure}[h]
 \centering
   \includegraphics[width=0.8\textwidth]{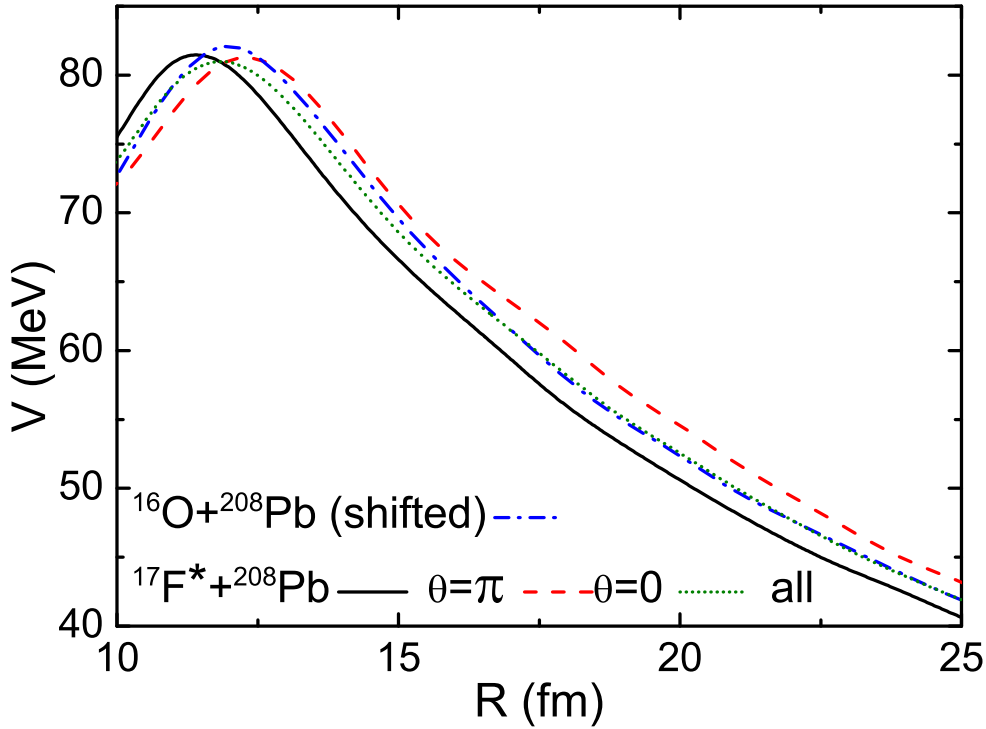}
   \caption{\label{fig7}(color online) Effective interaction potentials for fusion reaction \F17e+$^{208}$Pb at the energy $E_{\rm c.m.}=$ 85 MeV and $b=0$ fm, and $^{16}$O+$^{208}$Pb at the energy $E_{\rm c.m.}=$ 76 MeV and $b=0$ fm.}
   \end{figure}

   As well known, the classical fusion cross section $\sigma_{\rm fus}=\pi R_{\rm eff}^2$, where $R_{\rm eff}$ is the effective interaction radius directly relative to the radii of projectile and target. So the system with larger projectile and target nuclei will give larger fusion cross section. The rms radii for $^{17}$F and $^{16}$O are similar for the ground state, once $^{17}$F is excited by $^{208}$Pb to be a proton-halo nucleus with larger radius, the fusion cross section should be enhanced compared to $^{16}$O+$^{208}$Pb. But the enhancement is not observed in the experiments. Taken the shielding effect during the reactions into consideration, when the valence proton is pushed toward or even behind $^{16}$O core, the thickness of side of halo facing the target becomes small. That will reduce the effective interaction radius and change the dynamic effective interaction potential between projectile and target nuclei. In figure \ref{fig7}, we show the dynamic effective interaction potential for fusion reaction \F17e+$^{208}$Pb at the energy $E_{\rm c.m.}=$ 85 MeV and $b=0$ fm for valence proton distributing uniformly and for the two extreme case, i.e., the valence proton locating in the front of $^{16}$O core ($\theta=0$) and being totally shielded ($\theta=\pi$), respectively. One can see that, before projectile touch target, the effective interaction potential for the case of $\theta=0$ is larger than one of $\theta=\pi$; While after projectile touch target, the effective interaction potential for the case of $\theta=\pi$ is larger than one of $\theta=0$. The competition between the case for valence proton being shielded or not is one of possible reasons leading to no expected enhancement of the fusion cross section being observed. The result of $^{16}$O+$^{208}$Pb at the energy $E_{\rm c.m.}=$ 76 MeV and $b=0$ fm is also shown in the figure. Once again, the result of $^{16}$O+$^{208}$Pb is shifted in energy by factor 9/8. The similarity of the effective interaction potentials of \F17e+$^{208}$Pb and $^{16}$O+$^{208}$Pb indicates that these two systems have essentially the similar behavior, such as close fusion excitation functions.

\section{Summary}

 The experiments indicate that fusion cross sections of proton-halo nucleus is not enhanced comparing to the regular nucleus as people's expectation. One reasonable supposition is that the valence proton moves away from the target nucleus and is shielded by the core of halo nucleus, which is so called shielding effect. This effect has been verified with the fusion reactions by using the ImQMD model in this work. To obtain the good initial proton-halo nucleus, the special treatment on sample method for initial nucleus is firstly introduced into the ImQMD model. The shielding phenomenon is verified in the reactions \F17e+$^{208}$Pb by tracing the relative position of the valence proton to the core $^{16}$O with the microscopic dynamics simulations. The probability of valence proton being shielded increases when projectile and target get more and more close to each other. And it is found that before projectile and target touch each other, shielding effect reduces the effective interaction potential. While after projectile and target touch each other, shielding effect increases the effective interaction potential. The similar effective interaction potentials for $^{17}$F+$^{208}$Pb and $^{16}$O+$^{208}$Pb lead to similar fusion excitation functions for these two systems.

\section{Acknowledgement}
 The authors thank Dr. Ning Wang for careful reading of the manuscript and the valuable discussions.
 This work has been supported by
 National Natural Science Foundation of China
 (11965004, %Ou
 U1867212, %Wang
 11711540016, %NSFC-JSPS Zhou
 11847317), %Liang
 Natural Science Foundation of Guangxi province (2016GXNSFFA380001, 2017GXNSFGA198001),
 Foundation of Guangxi innovative team and distinguished scholar in institutions of higher education, and
 Innovation project of Guangxi Graduate Education (XYCSZ2018051).

\clearpage

%\end{CJK*}
\end{document}